\documentclass[conference]{IEEEtran}
\IEEEoverridecommandlockouts
\usepackage{cite}
\usepackage{amsmath,amssymb,amsfonts}
\usepackage{algorithmic}
\usepackage{graphicx}
\usepackage{textcomp}
\usepackage{xcolor}
\usepackage[hyphens]{url}  % 允许在连字符处断行
\usepackage{hyperref}
\usepackage{tabularx}   % For auto-wrapping text
\usepackage{booktabs}
\usepackage{graphicx} % 用于 resizebox
\usepackage{multirow}
\usepackage{amssymb}   % 用于 \checkmark 和 \times

\usepackage{colortbl}   % 单元格背景色
\usepackage{xcolor}     % 颜色处理

\definecolor{bg_gray}{gray}{0.92}
\newcommand{\g}[1]{\textcolor{gray}{#1}}

\def\BibTeX{{\rm B\kern-.05em{\sc i\kern-.025em b}\kern-.08em
    T\kern-.1667em\lower.7ex\hbox{E}\kern-.125emX}}
\begin{document}

\title{Integrating Fine-Grained Audio-Visual Evidence for Robust Multimodal Emotion Reasoning}

% \author{
%     % --- 作者姓名部分 ---
%     \IEEEauthorblockN{
%         Zhixian Zhao\IEEEauthorrefmark{2}\IEEEauthorrefmark{3},
%         Wenjie Tian\IEEEauthorrefmark{2}\IEEEauthorrefmark{3},
%         Xiaohai Tian,
%         Jun Zhang and
%         Lei Xie\IEEEauthorrefmark{1}\IEEEauthorrefmark{3},~\IEEEmembership{Senior Member,~IEEE}
%         % --- 关键点：将 thanks 放在姓名块的最后 ---
%         \thanks{\IEEEauthorrefmark{2}Equal contribution (email: zxzhao, twj@mail.nwpu.edu.cn).}
%         \thanks{\IEEEauthorrefmark{1}Corresponding author (email: lxie@nwpu.edu.cn).}
%     }
    
%     % --- 机构部分 ---
%     % 假设 Lei Xie 是机构 1 (NWPU)
%     \IEEEauthorblockA{\IEEEauthorrefmark{3}Audio, Speech and Language Processing Group (ASLP@NPU), \\
%     Northwestern Polytechnical University, Xi'an, China}
% }

\author{
    % --- 作者姓名部分 ---
    \IEEEauthorblockN{
        Zhixian Zhao\IEEEauthorrefmark{2}\IEEEauthorrefmark{3},
        Wenjie Tian\IEEEauthorrefmark{2}\IEEEauthorrefmark{3} and
        Lei Xie\IEEEauthorrefmark{1}\IEEEauthorrefmark{3},~\IEEEmembership{Senior Member,~IEEE}
        % --- 关键点：将 thanks 放在姓名块的最后 ---
        \thanks{\IEEEauthorrefmark{2}Equal contribution (email: zxzhao, twj@mail.nwpu.edu.cn).}
        \thanks{\IEEEauthorrefmark{1}Corresponding author (email: lxie@nwpu.edu.cn).}
    }
    
    % --- 机构部分 ---
    % 假设 Lei Xie 是机构 1 (NWPU)
    \IEEEauthorblockA{\IEEEauthorrefmark{3}Audio, Speech and Language Processing Group (ASLP@NPU), \\
    Northwestern Polytechnical University, Xi'an, China}
}

\maketitle

\begin{abstract}

Multimodal emotion analysis is shifting from static classification to generative reasoning. 
Beyond simple label prediction, robust affective reasoning must synthesize fine-grained signals such as facial micro-expressions and prosodic which shifts to decode the latent causality within complex social contexts.
% 缺点不够，数据：第一个贡献点：提出数据集。第二个缺点：多模态信息并没有融合的很好。第二个贡献点：Structured Evidence Decomposition paradigm and CA-DPO
However, current Multimodal Large Language Models (MLLMs) face significant limitations in fine-grained perception, primarily due to data scarcity and insufficient cross-modal fusion. 
As a result, these models often exhibit unimodal dominance which leads to hallucinations in complex multimodal interactions, particularly when visual and acoustic cues are subtle, ambiguous, or even contradictory (e.g., in sarcastic scenery).
To address this, we introduce SABER-LLM, a framework designed for robust multimodal reasoning. 
First, we construct SABER, a large-scale emotion reasoning dataset comprising 600K video clips, annotated with a novel six-dimensional schema that jointly captures audiovisual cues and causal logic. 
Second, we propose the structured evidence decomposition paradigm, which enforces a “perceive-then-reason” separation between evidence extraction and reasoning to alleviate unimodal dominance.
% 增强complex 场景的 perceive
The ability to perceive complex scenes is further reinforced by consistency-aware direct preference optimization, which explicitly encourages alignment among modalities under ambiguous or conflicting perceptual conditions.
Experiments on EMER, EmoBench-M, and SABER-Test demonstrate that SABER-LLM significantly outperforms open-source baselines and achieves robustness competitive with closed-source models in decoding complex emotional dynamics. The dataset and model are available at \url{https://github.com/zxzhao0/SABER-LLM}.
\end{abstract}

\begin{IEEEkeywords}
Multimodal Emotion Reasoning, Multimodal Large Language Model, Fine-grained Perception, Dataset Construction
\end{IEEEkeywords}

\vspace{-5pt}
\section{Introduction}
\label{sec:intro}

While Multimodal Large Language Models (MLLMs)~\cite{llava,qwen3-omni,gpt4o} have advanced generative intelligence, their potential in fine-grained emotion reasoning remains largely untapped. Unlike conventional approaches that assign discrete labels, generative reasoning requires synthesizing multimodal signals—ranging from acoustic prosody to micro-expressions to interpret the causal logic behind complex social interactions.

\begin{figure}[t!]
  \centering
  \includegraphics[width=1\linewidth]{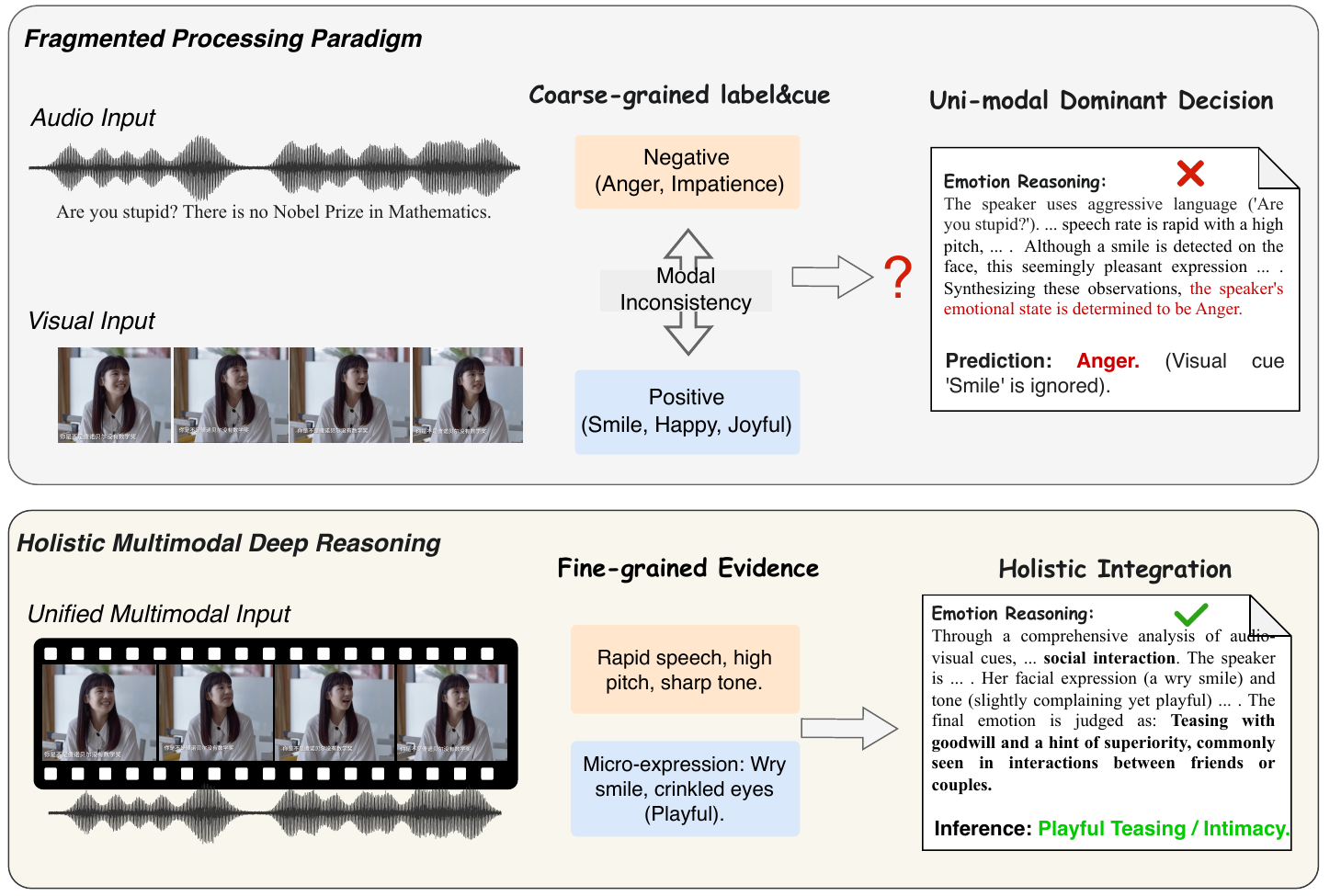}
  \caption{Fragmented vs. Holistic Emotion Reasoning. Top: Conventional methods suffer from uni-modal dominance, failing to reconcile contradictory cues (e.g., a smile masking a sharp tone) and misinterpreting the intent. Bottom: Holistic reasoning integrates fine-grained evidence to correctly decipher complex social dynamics (e.g., ``Playful Teasing")}
  \label{overview}
\vspace{-13pt}
\end{figure}

However, existing systems suffer from insufficient fine-grained perception. As illustrated in the upper part of Figure~\ref{overview}, current frameworks~\cite{mosear,emochat,emotion-llama} typically operate in a ``fragmented'' paradigm, independently processing visual and acoustic signals to predict static labels. This isolation makes models susceptible to unimodal dominance, where they over-rely on a single salient modality while neglecting subtle yet critical evidence from other channels. 
Meanwhile, this fragmented paradigm also overlooks cross-modal context, rendering models susceptible to critical misjudgments when cues are contradictory. For instance, capturing a ``smile'' while missing an ``impatient tone'' may result in misinterpreting ``playful teasing'' as genuine conflict. 
This limitation not only causes failure in modality-inconsistent scenarios but also degrades performance in tasks requiring nuance detection (e.g., a ``polite but cold tone''). We attribute these limitations primarily to two factors: the scarcity of fine-grained, structured data that teaches models ``why'' an emotion occurs, and the lack of mechanisms to enforce explicit perception, allowing models to bypass evidence grounding in favor of statistical shortcuts.

To address these challenges, we introduce SABER-LLM. First, we establish SABER (Scene, Audio, Body, Expression, and Reasoning), a large-scale dataset comprising approximately 600k video clips. SABER features a unique six-dimensional annotation schema that covers features spanning from macro scenes to micro-expressions. Second, to mitigate uni-modal dominance, we propose the Structured Evidence Decomposition (SED) paradigm. 
Diverging from standard instruction tuning, SED compels the model to independently disentangle and analyze uni-modal evidence before synthesizing a final conclusion. Furthermore, we incorporate a consistency-aware direct preference optimization (CA-DPO) strategy to align the model's judgment with human standards when resolving audio-visual discrepancies.

In summary, our main contributions are:
\begin{itemize}
    \item We introduce SABER, a large-scale structured multimodal emotion reasoning dataset containing approximately 600k clips. It is constructed via a scalable pipeline that integrates a unified fine-grained annotation strategy with automated quality control mechanisms. Additionally, we release SABER-Test to evaluate model robustness across modalities.
    \item We propose the SABER-LLM framework, incorporating the structured evidence decomposition paradigm and CA-DPO. By explicitly decoupling evidence perception from the reasoning process, our approach effectively mitigates uni-modal dominance, achieving significant performance improvements in both general emotion understanding and complex modality-inconsistent scenarios.
    \item Extensive experiments demonstrate that our 7B model outperforms existing open-source baselines on public benchmarks (EMER, EmoBench-M) and SABER-Test, exhibiting competitive performance against closed-source models like Gemini-2.5-Pro across multiple metrics.
\end{itemize}

\vspace{-5pt}
\section{Related Work}
\vspace{-5pt}
\subsection{Multimodal Large Language Models}

Multimodal Large Language Models (MLLMs) align visual and auditory modalities with the semantic space of LLMs. Early research primarily targeted static image understanding, where approaches like LLaVA~\cite{llava} and Qwen-VL~\cite{qwen2.5vl} used projection layers to map visual features into text embeddings. This paradigm was later extended to dynamic video and audio domains. Models such as Video-LLaMA~\cite{VideoLLaMA} and Qwen-Audio~\cite{qwen-audio} introduced temporal aggregation and auditory encoders, enabling the comprehension of temporal evolution and non-speech acoustic events.

Recently, the field has evolved towards ``Omni-modal'' interaction. Native multimodal models, including Qwen-Omni series~\cite{qwen3-omni,Qwen2.5-Omni} and GPT-4o~\cite{gpt4o}, exhibit strong capabilities in processing interleaved video, audio, and text streams. Despite their success in general-purpose tasks, these MLLMs often lack the fine-grained domain perception necessary for affective computing. As noted in~\cite{mosear,emochat}, they frequently miss subtle cues in micro-expressions or prosody, leading to poor performance in complex affective reasoning.

\vspace{-5pt}
\subsection{Multimodal Emotion Reasoning}
\vspace{-5pt}
Multimodal emotion analysis interprets human emotional states from heterogeneous signals. While traditional discriminative approaches perform robustly on emotion recognition benchmarks, such as CREMA-D~\cite{CREMA-D}, MELD~\cite{meld}, they cannot provide causal explanations or handle open-vocabulary scenarios.

MLLMs have facilitated the shift to multimodal emotion reasoning. Emotion-LLaMA~\cite{emotion-llama} pioneered reasoning via instruction tuning. To capture finer nuances, EmoChat~\cite{emochat} explicitly modeled facial Action Units. Addressing semantic conflicts, MOSEAR~\cite{mosear} identified ``Audio Bias'' in sarcastic scenarios and proposed attention reallocation strategies. However, these methods rely on implicit feature fusion within the LLM, making them susceptible to uni-modal dominance and hallucinations when facing contradictory audio-visual cues (e.g., ``smiling with a knife''). 
We address this limitation with SABER-LLM. By employing a SED strategy, we enforce an explicit ``perceive-then-reason'' process, forcing the model to base its conclusions on verified unimodal evidence.
% \begin{table}[t]
% \centering
% \caption{Overview of Collected Multimodal Datasets.}
% \label{tab:datasets}
% % \footnotesize % 稍微缩小字体以适应单栏
% % \setlength{\tabcolsep}{3.5pt} % 减小列间距
% \begin{tabular}{l c c c}
% \toprule
% \textbf{Dataset} & \textbf{Lang} & \textbf{Scale (\#Utt)} & \textbf{Source} \\
% \midrule
% CREMA-D~\cite{} & EN & 7.4k & Acted \\
% MEAD~\cite{} & EN & 31.7k & Acted \\
% MELD~\cite{} & EN & 13.7k & TV \\
% MEIJU25 (CN)~\cite{} & CN & 114.1k & TV \\
% MEIJU25 (EN)~\cite{} & EN & 85.9k & TV \\
% MER25~\cite{} & CN & 132.2k & TV \\
% MSP-IMPROV~\cite{} & EN & 1.3k & Acted \\
% MultiDialog~\cite{} & EN & 187.9k & Acted \\
% RAVDESS~\cite{} & EN & 2.9k & Acted \\
% CH-SIMSv2.0-s~\cite{} & CN & 4.4k & Movie/TV \\
% MEmoR~\cite{} & EN & 22.7k & TV \\
% \bottomrule
% \end{tabular}
% \vspace{-10pt} % 调整表后间距
% \end{table}

\begin{table}[t]
\centering
\caption{Overview of Collected Multimodal Datasets. Unlike existing datasets that focus on \textbf{Recognition (Rec.)}, SABER is uniquely annotated for \textbf{Fine-grained Reasoning}. It provides not only holistic analysis ($R_{All}$) but also decoupled evidence for Audio ($R_A$) and Visual ($R_V$) modalities.}
\label{tab:datasets}
\setlength{\tabcolsep}{1.5pt} % 压缩列间距
% 修复点1：确保 resizebox 的大括号在 end{tabular} 后闭合
\resizebox{\linewidth}{!}{
    \begin{tabular}{l c c c c c c c c}
    \toprule
    \multirow{2}{*}{\textbf{Dataset}} & \multirow{2}{*}{\textbf{Lang}} & \multirow{2}{*}{\textbf{Mod.}} & \multirow{2}{*}{\textbf{Scale}} & \multirow{2}{*}{\textbf{Source}} & \textbf{Rec.} & \multicolumn{3}{c}{\textbf{Reasoning}} \\
    \cmidrule(lr){6-6} \cmidrule(l){7-9} 
     & & & & & \textbf{Emo.} & \textbf{$R_A$} & \textbf{$R_V$} & \textbf{$R_{All}$} \\
    \midrule
    % 修复点2：给 \times 和 \checkmark 加上 $ 符号，使其在数学模式下运行
    CREMA-D~\cite{CREMA-D} & EN & V+A & 7.4k & Acted & $\checkmark$ & $\times$ & $\times$ & $\times$ \\
    MEAD~\cite{mead} & EN & V+A & 31.7k & Acted & $\checkmark$ & $\times$ & $\times$ & $\times$ \\
    MELD~\cite{meld} & EN & V+A & 13.7k & TV & $\checkmark$ & $\times$ & $\times$ & $\times$ \\
    MEIJU25 (CN)~\cite{meiju} & CN & V+A & 114.1k & TV & $\checkmark$ & $\times$ & $\times$ & $\times$ \\
    MEIJU25 (EN)~\cite{meiju} & EN & V+A & 85.9k & TV & $\checkmark$ & $\times$ & $\times$ & $\times$ \\
    MER25~\cite{mer25} & CN & V+A & 132.2k & TV & $\checkmark$ & $\times$ & $\times$ & $\times$ \\
    MSP-IMPROV~\cite{msp-improv} & EN & V+A & 1.3k & Acted & $\checkmark$ & $\times$ & $\times$ & $\times$ \\
    MultiDialog~\cite{multidialog} & EN & V+A & 187.9k & Acted & $\checkmark$ & $\times$ & $\times$ & $\times$ \\
    RAVDESS~\cite{ravdess} & EN & V+A & 2.9k & Acted & $\checkmark$ & $\times$ & $\times$ & $\times$ \\
    CH-SIMSv2.0-s~\cite{chsimsv2} & CN & V+A & 4.4k & Movie/TV & $\checkmark$ & $\times$ & $\times$ & $\times$ \\
    MEmoR~\cite{MEmoR} & EN & V+A & 22.7k & TV & $\checkmark$ & $\times$ & $\times$ & $\times$ \\
    \midrule
    \textbf{SABER (Ours)} & \textbf{CN/EN} & \textbf{V+A} & \textbf{600k} & \textbf{Mixed} & \textbf{$\checkmark$} & \textbf{$\checkmark$} & \textbf{$\checkmark$} & \textbf{$\checkmark$} \\
    \bottomrule
    \end{tabular}
} % <--- 这一行的括号是必须的，用于闭合 resizebox
\vspace{-10pt}
\end{table}

\begin{figure*}[t!]
  \centering
  \includegraphics[width=0.85\linewidth]{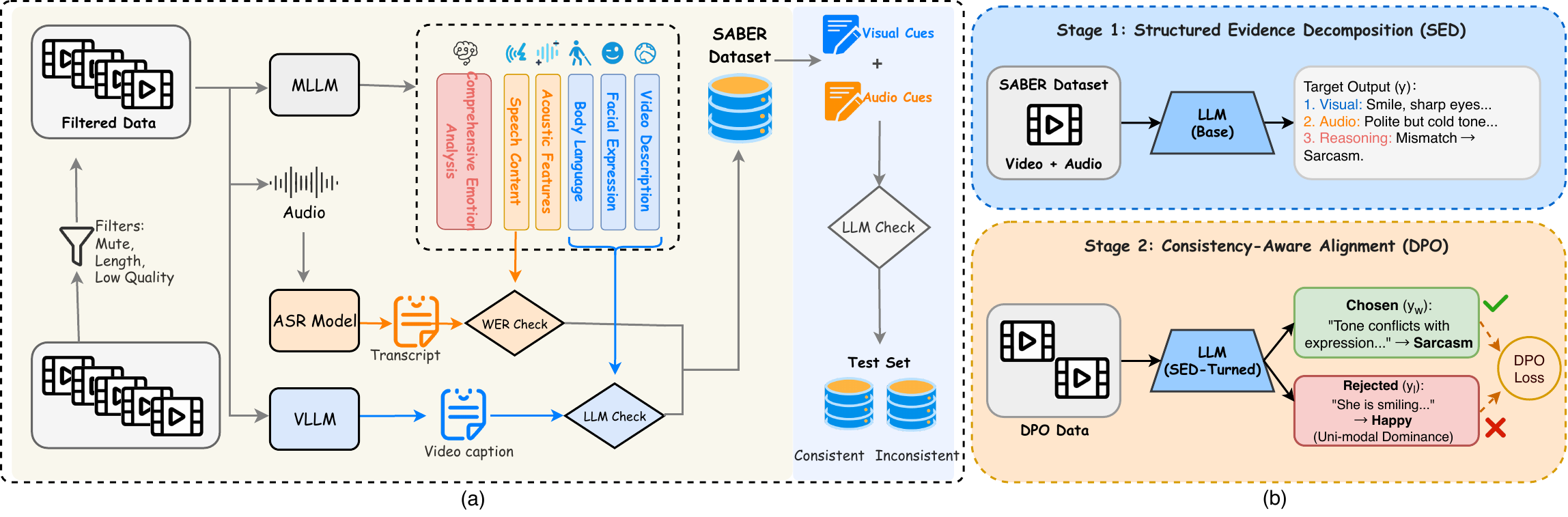}  % 使用单栏宽度的80%
\caption{
(a) Data Pipeline: A scalable construction process featuring six-dimensional fine-grained annotation and automated hallucination filtering. (b) Training Paradigm: The Structured Evidence Decomposition (SED) stage (Stage 1) enforces sequential "perceive-then-reason" grounding, followed by CA-DPO (Stage 2) to align reasoning with human preference in modality-conflicting scenarios.}
  \label{fig:dataset_model}
\vspace{-10pt}
\end{figure*}

\vspace{-5pt}
\section{SABER Dataset}
\label{sec:dataset}
\vspace{-5pt}
To bridge the gap between superficial classification and deep affective reasoning, we introduce \textbf{SABER}, a dataset featuring a novel six-dimensional fine-grained annotation schema to ground reasoning in observable multimodal evidence. The construction pipeline is illustrated in Figure~\ref{fig:dataset_model}a.

% \paragraph{Data Construction Pipeline}
% We aggregated diverse data from multiple sources, as presented in Table~\ref{tab:datasets}, with a total of 600k clips. To ensure data quality, samples were filtered based on three criteria: technical anomalies, excessive background noise, and abnormal durations. Technical anomalies include low resolution and missing tracks, while abnormal durations refer to those shorter than 0.5 seconds or longer than 30 seconds.

\paragraph{Data Construction Pipeline}
We aggregated diverse data from multiple sources, as presented in Table~\ref{tab:datasets}, with a total of 600k clips. To ensure data quality, samples were filtered based on three criteria: technical anomalies, excessive background noise, and abnormal durations. Technical anomalies include low resolution and missing tracks, while abnormal durations refer to those shorter than 0.5 seconds or longer than 30 seconds.

% Since source datasets lack causal reasoning information, we employed a Unified Annotation Strategy using Gemini-2.5-Pro~\cite{gemini2.5} with synchronized audio-visual inputs. We designed a hierarchical prompting scheme covering six dimensions: \textbf{(1) Video description} (macro-scene context); \textbf{(2) Speech content} (verbatim transcripts and semantics); \textbf{(3) Acoustic features} (prosody, pitch, and tonal intensity, e.g., ``trembling and rapid''); \textbf{(4) Facial expression} (micro-expressions, gaze); \textbf{(5) Body language} (posture, gestures, and social signals); and \textbf{(6) Comprehensive reasoning}, synthesizing evidence to deduce the causal logic behind the emotion.
Given the lack of causal reasoning information in the source datasets, a unified annotation strategy was adopted using Gemini-2.5-Pro~\cite{gemini2.5} with synchronized audio-visual inputs. A hierarchical prompting scheme was designed to cover six dimensions, namely video description, speech content, acoustic features, facial expression, body language and comprehensive reasoning. Specifically, video description focuses on macro-scene context; speech content involves verbatim transcripts and semantic information; acoustic features include prosody, pitch and tonal intensity; facial expression covers micro-expressions and gaze; body language encompasses posture, gestures and social signals; comprehensive reasoning synthesizes multi-dimensional evidence to deduce the causal logic underlying emotions.

\paragraph{Quality Assurance and Diagnostic Benchmark}
As shown in Figure~\ref{fig:dataset_model}a, we implement a two-step verification to mitigate hallucinations. First, we use an ASR model to obtain ground-truth transcripts and filter auditory hallucinations via Word Error Rate (WER) checks against Gemini-generated content. Second, we employ Qwen2.5-VL to generate independent video descriptions and utilize an LLM to assess semantic consistency with SABER annotations, discarding samples with significant deviations.
To explicitly probe robustness against modality conflicts, we construct the SABER-Test set and design a two-stage consistency assessment protocol. We first employ GPT-4o~\cite{gpt4o} to classify the relationship between acoustic and visual cues as consistency or inconsistency. Secondly, we assign a quantitative intensity score ranging from 0 to 10, which represents the degree of alignment or conflict severity.
Based on this metric, we selected 1,800 top-ranked samples, evenly divided into a consistent subset and an inconsistent subset, providing a precise testbed for diagnosing uni-modal dominance.

\vspace{-5pt}
\section{Method}
\label{sec:method}
\vspace{-5pt}
We introduce SABER-LLM, a robust framework for multimodal emotion reasoning built upon the Qwen2.5-Omni architecture~\cite{Qwen2.5-Omni}. As illustrated in Figure~\ref{fig:dataset_model}b, our training pipeline operates in two stages: (1) Structured Evidence Decomposition (SED), a reasoning paradigm implemented via supervised fine-tuning to inject a ``perceive-then-reason'' inductive bias; and (2) Consistency-Aware Preference Alignment, which employs CA-DPO to refine the model's sensitivity to cross-modal conflicts.
\vspace{-5pt}
\subsection{Structured Evidence Decomposition}
\label{subsec:sed}
\vspace{-5pt}
Unconstrained end-to-end paradigms typically map multimodal inputs directly to a final conclusion. This often allows models to exploit superficial uni-modal shortcuts (e.g., focusing solely on a smile) rather than synthesizing heterogeneous signals, exacerbating the uni-modal dominance problem.

To address this, we model emotion understanding as a serialized, evidence-based reasoning process. Let $\mathcal{V}$ and $\mathcal{A}$ denote the visual and acoustic inputs, respectively, forming the multimodal input $X = \{\mathcal{V}, \mathcal{A}\}$. We define the target generation sequence as a structured triplet $Y = [E_{v}, E_{a}, R]$, where $E_{v}$ denotes visual evidence (e.g., facial micro-expressions, scene context), $E_{a}$ represents acoustic evidence (e.g., prosody, tone intensity), and $R$ stands for holistic reasoning, which synthesizes the above evidence to deduce causal logic.

Instead of directly mapping inputs to reasoning, we impose a structural constraint (Eq.~\ref{eq:sed_prob}) that compels the model to first explicitly extract acoustic evidence ($E_a$) and visual evidence ($E_v$) before formulating the holistic reasoning ($R$).
\vspace{-2pt}
\begin{equation}
\footnotesize
P(Y | X, I) = \underbrace{P(E_{v} | X, I)}_{\text{Visual}} \cdot \underbrace{P(E_{a} | E_{v}, X, I)}_{\text{Acoustic}} \cdot \underbrace{P(R | E_{v}, E_{a}, X, I)}_{\text{Reasoning}},
\label{eq:sed_prob}
\end{equation}
\vspace{-2pt}
where $I$ denotes the system instruction. We order visual extraction first to establish a scene baseline, guiding the model to explicitly target fine-grained acoustic nuances conditioned on the visual context.

This decomposition introduces a strong inductive bias: the model is explicitly supervised to disentangle unimodal evidence before performing high-level reasoning. This effectively prevents the model from skipping the perception stage. We optimize the model parameters $\theta$ by minimizing the negative log-likelihood on the SABER dataset $\mathcal{D}_{\text{SABER}}$:
\begin{equation}
\mathcal{L}_{\text{SFT}} = - \mathbb{E}_{(X, Y) \sim \mathcal{D}_{\text{SABER}}} \left[ \sum_{t=1}^{|Y|} \log P_\theta(y_t | y_{<t}, X, I) \right].
\label{eq:sft_loss}
\end{equation}
By enforcing this structured generation path, the model learns to ground its final predictions in specific auditory and visual observations, significantly reducing hallucinations.
\vspace{-5pt}
\subsection{Consistency-Aware Preference Alignment}
\label{subsec:dpo}
\vspace{-2pt}
While the stage-1 model ($\pi_{\text{sft}}$) adopts the structured reasoning format, it remains susceptible to internal logical inconsistencies. Supervised fine-tuning(SFT) often prioritizes sequence-level imitation over logical grounding, leading the model to identify conflicting cues (e.g., a ``polite smile'' vs. a ``hostile tone'') yet revert to a biased inference that aligns with only the most salient modality. 
To resolve these contradictions, we introduce CA-DPO. Unlike SFT, our framework explicitly contrasts logically grounded reasoning trajectories against internally contradictory ones during preference learning. 
Specifically, CA-DPO penalizes output sequences that neglect or override previously extracted unimodal evidence, thereby enforcing consistency between perception and reasoning stages. As a result, the “perceive-then-reason” pipeline transitions from a superficial structural template into a verifiable chain of evidence-based deduction.

%Although the Stage-1 model ($\pi_{\text{sft}}$) learns structured reasoning, it remains susceptible to cross-modal logical fallacies (e.g., hallucinating a happy tone to match a smile). SFT mimics data distribution but lacks the discriminative power to distinguish grounded truth from plausible hallucinations. To bridge this gap, we employ Direct Preference Optimization (DPO). Unlike PPO-based RLHF which requires a separate reward model and is prone to instability, DPO directly optimizes the policy on preference data, offering a more stable and efficient alignment process for fine-grained multimodal constraints.

As shown in Stage 2 of Figure~\ref{fig:dataset_model}b, we construct a preference dataset $\mathcal{D}_{\text{pref}}$ focused on modality consistency. 
We sample a balanced subset from the training data containing both consistent and inconsistent emotional samples. For each input $X$, we generate $K$ candidate responses $\{r_1, ..., r_K\}$ using $\pi_{\text{sft}}$. 
To ensure high-quality preference labels, we employ GPT-4o as an evaluator to score these candidates based on their logical consistency and evidence alignment. 
% Based on the rankings, we select the top-2 highest-scoring responses as positive samples and the bottom-2 lowest-scoring ones as negative samples to form multiple pairs. 
Based on these scores, we rank the candidates and form preference pairs by selecting the two highest-ranked responses as winner~$y_w$ and the two lowest-ranked as loser~$y_l$.
% We apply a consistency-aware screening strategy to categorize these: winner ($y_w$) responses correctly identify and reconcile the relationship between acoustic and visual cues (e.g., recognizing a smile masking an angry tone), whereas loser ($y_l$) responses fail to detect discrepancies, suffer from uni-modal dominance, or hallucinate non-existent cues.
As a result, winner responses correctly identify and reconcile the relationship between acoustic and visual cues whereas loser responses fail to detect discrepancies, suffer from uni-modal dominance, or hallucinate non-existent cues.

We optimize the policy $\pi_\theta$ to maximize the margin between the winner consistent path $y_w$ and the loser flawed path $y_l$. Setting the frozen SFT model $\pi_{\text{sft}}$ as the reference model $\pi_{\text{ref}}$, the CA-DPO objective is formalized as:
\begin{equation}
\small
\begin{split}
\mathcal{L}_{\text{DPO}} = - \mathbb{E}_{(X, y_w, y_l) \sim \mathcal{D}_{\text{pref}}} \Big[ \log \sigma \Big( & \beta \log \frac{\pi_\theta(y_w|X)}{\pi_{\text{ref}}(y_w|X)} \\
& - \beta \log \frac{\pi_\theta(y_l|X)}{\pi_{\text{ref}}(y_l|X)} \Big) \Big].
\end{split}
\label{eq:dpo_loss}
\end{equation}
Here, $\sigma$ is the sigmoid function and $\beta$ controls the deviation penalty. This objective forces the model to prioritize evidence consistency over superficial statistical correlations, achieving understanding of complex emotional dynamics.

\vspace{-5pt}
\section{Experiments}

\begin{table*}[t]
\centering
\caption{Performance comparison on EmoBench-M. ``Avg.'' columns indicate the average accuracy within each category. The best results for open-source models are bolded, and the second-best are \underline{underlined}. Task abbreviations are detailed in Sec.~\ref{sec:setup}.}
\label{tab:emobench}
% 自动缩放表格以适应页面宽度
\resizebox{\textwidth}{!}{
\begin{tabular}{l ccccc c ccccc c ccc c}
\toprule
% 表头
\textbf{Model} & SOER & SPER & OSA & EIA & SCEA & \cellcolor{bg_gray}Avg.(FER) & FGDEA & PEA & FCDEA & CEIA & MPDER & \cellcolor{bg_gray}Avg.(CEU) & HU & SD & \cellcolor{bg_gray}Avg.(SCEA) & \cellcolor{bg_gray}all\_Avg. \\
\midrule

% --- 开源模型 ---
\multicolumn{17}{l}{\textit{Open-source Models}} \\ 
\midrule
Intern-S1-9B~\cite{intern-s1} & 39.96 & 34.91 & 62.35 & 53.68 & 40.25 & \cellcolor{bg_gray}46.23 & 53.35 & \underline{69.49} & 66.96 & 6.67 & 37.72 & \cellcolor{bg_gray}\underline{46.84} & 49.18 & 65.76 & \cellcolor{bg_gray}57.47 & \cellcolor{bg_gray}50.18 \\
Video-LLaMA2.1-7B~\cite{VideoLLaMA} & 50.40 & 37.70 & \textbf{73.00} & 57.60 & 33.20 & \cellcolor{bg_gray}50.40 & 51.50 & 68.20 & \underline{67.60} & 6.50 & 36.60 & \cellcolor{bg_gray}46.10 & 54.70 & 53.40 & \cellcolor{bg_gray}54.10 & \cellcolor{bg_gray}48.70 \\
LongVA-DPO-7B~\cite{longva} & 50.20 & 44.20 & 33.80 & 45.70 & \textbf{54.80} & \cellcolor{bg_gray}45.70 & 51.10 & 33.20 & 33.30 & 6.10 & 37.00 & \cellcolor{bg_gray}32.10 & 63.60 & 51.60 & \cellcolor{bg_gray}57.60 & \cellcolor{bg_gray}43.80 \\
InternVideo2-Chat-8B~\cite{InternVideo2} & 55.20 & 44.00 & 45.40 & 56.00 & \underline{52.40} & \cellcolor{bg_gray}50.60 & 58.00 & 50.80 & 49.20 & 8.90 & 34.20 & \cellcolor{bg_gray}40.20 & \underline{68.10} & 61.20 & \cellcolor{bg_gray}\underline{64.70} & \cellcolor{bg_gray}51.50 \\
MiniCPM-V-2.6-8B~\cite{minicpm} & 26.60 & 21.80 & 56.50 & 50.50 & 44.50 & \cellcolor{bg_gray}40.00 & 48.90 & 58.60 & 57.10 & 11.70 & 39.20 & \cellcolor{bg_gray}43.10 & 55.10 & 49.60 & \cellcolor{bg_gray}52.40 & \cellcolor{bg_gray}46.50 \\
InternVL2.5-8B~\cite{internvl} & 40.30 & 40.80 & \underline{67.80} & \textbf{62.00} & 45.00 & \cellcolor{bg_gray}\underline{51.20} & 48.90 & 61.00 & 62.50 & \textbf{12.40} & \textbf{43.80} & \cellcolor{bg_gray}45.70 & 66.50 & 59.60 & \cellcolor{bg_gray}63.10 & \cellcolor{bg_gray}50.40 \\
Emotion-LLaMA-7B~\cite{emotion-llama} & 44.80 & 33.40 & 23.00 & 41.10 & 42.00 & \cellcolor{bg_gray}36.90 & \underline{62.00} & 24.60 & 25.20 & 2.90 & 38.90 & \cellcolor{bg_gray}30.70 & 58.00 & 53.00 & \cellcolor{bg_gray}55.50 & \cellcolor{bg_gray}40.60 \\
Qwen2.5-Omni-7B~\cite{Qwen2.5-Omni} & \textbf{58.80} & \underline{53.80} & 59.80 & 12.27 & 51.60 & \cellcolor{bg_gray}47.25 & 60.10 & 58.72 & 59.74 & 7.01 & \underline{40.80} & \cellcolor{bg_gray}45.27 & 62.50 & \underline{65.80} & \cellcolor{bg_gray}64.15 & \cellcolor{bg_gray}\underline{52.23} \\
SABER-LLM & \underline{58.00} & \textbf{62.00} & 65.80 & \underline{61.37} & 50.80 & \cellcolor{bg_gray}\textbf{59.59} & \textbf{62.04} & \textbf{76.15} & \textbf{73.96} & \underline{12.00} & 36.60 & \cellcolor{bg_gray}\textbf{52.15} & \textbf{69.42} & \textbf{66.40} & \cellcolor{bg_gray}\textbf{67.91} & \cellcolor{bg_gray}\textbf{59.88} \\

\midrule
% --- 大参数模型 (全灰) ---
\multicolumn{17}{l}{\textit{Large-scale Models}} \\
\midrule
\g{Qwen3-Omni-30B~\cite{qwen3-omni}} & \g{46.40} & \g{42.40} & \g{71.00} & \g{54.33} & \g{53.20} & \cellcolor{bg_gray}\g{53.47} & \g{76.40} & \g{69.54} & \g{68.27} & \g{10.00} & \g{51.00} & \cellcolor{bg_gray}\g{55.04} & \g{76.34} & \g{64.80} & \cellcolor{bg_gray}\g{70.57} & \cellcolor{bg_gray}\g{59.69} \\
\g{InternVL2.5-78B~\cite{internvl}} & \g{48.80} & \g{41.20} & \g{63.20} & \g{59.40} & \g{52.40} & \cellcolor{bg_gray}\g{53.00} & \g{52.70} & \g{56.80} & \g{56.70} & \g{12.60} & \g{43.50} & \cellcolor{bg_gray}\g{44.50} & \g{76.80} & \g{64.40} & \cellcolor{bg_gray}\g{70.60} & \cellcolor{bg_gray}\g{52.40} \\
\g{Qwen2.5-VL-72B~\cite{qwen2.5vl}} & \g{44.80} & \g{35.60} & \g{72.40} & \g{62.70} & \g{58.40} & \cellcolor{bg_gray}\g{53.00} & \g{51.60} & \g{64.20} & \g{64.30} & \g{11.40} & \g{47.80} & \cellcolor{bg_gray}\g{47.90} & \g{77.70} & \g{65.60} & \cellcolor{bg_gray}\g{71.70} & \cellcolor{bg_gray}\g{57.80} \\

\midrule
% --- 闭源模型 (全灰) ---
\multicolumn{17}{l}{\textit{Closed-source models}} \\
\midrule
\g{GLM-4V-PLUS~\cite{glm4}} & \g{54.90} & \g{43.70} & \g{70.00} & \g{61.20} & \g{50.80} & \cellcolor{bg_gray}\g{56.10} & \g{51.80} & \g{62.80} & \g{65.40} & \g{14.70} & \g{41.60} & \cellcolor{bg_gray}\g{47.30} & \g{74.70} & \g{59.80} & \cellcolor{bg_gray}\g{67.30} & \cellcolor{bg_gray}\g{57.70} \\
\g{Gemini-2.0-Flash~\cite{gemini2.0}} & \g{63.30} & \g{55.80} & \g{68.80} & \g{63.50} & \g{55.60} & \cellcolor{bg_gray}\g{61.40} & \g{64.20} & \g{70.90} & \g{71.90} & \g{11.10} & \g{48.70} & \cellcolor{bg_gray}\g{53.40} & \g{79.20} & \g{64.80} & \cellcolor{bg_gray}\g{72.00} & \cellcolor{bg_gray}\g{62.30} \\
\g{Gemini-2.5-Pro~\cite{gemini2.5}} & \g{70.40} & \g{60.20} & \g{75.20} & \g{60.36} & \g{59.60} & \cellcolor{bg_gray}\g{65.15} & \g{72.26} & \g{77.15} & \g{78.12} & \g{14.80} & \g{51.40} & \cellcolor{bg_gray}\g{58.75} & \g{79.02} & \g{72.40} & \cellcolor{bg_gray}\g{75.71} & \cellcolor{bg_gray}\g{66.54} \\
\bottomrule
\end{tabular}
}
\vspace{-12pt}
\end{table*}

\vspace{-5pt}
\subsection{Experimental Setup}
\label{sec:setup}
\noindent \textbf{Evaluation Benchmarks.} We evaluate our framework on three diverse benchmarks: 
(1) EMER~\cite{emer} (332 clips), assessing the depth of logical reasoning; 
(2) SABER-Test, our internal bilingual test set (1,800 clips) evenly split into ``consistent'' and ``inconsistent'' subsets to assess robustness against audio-visual conflicts; and 
(3) EmoBench-M, a comprehensive benchmark covering 13 tasks across three categories: \textit{Foundational Emotion Recognition}, including Speech Emotion Recognition (SPER), Emotion Intensity Analysis (EIA), Opinion Sentiment Analysis (OSA), Stock Comment Emotion Analysis (SCEA), and Song Emotion Recognition (SOER); \textit{Conversational Emotion Understanding}, including Fine-Grained Dialog Emotion Analysis (FGDEA), Conversational Emotion \& Intent Analysis (CEIA), Presentation Emotion Analysis (PEA), Face-Centric Dialog Emotion Analysis (FCDEA), and Multi-Party Dialog Emotion Recognition (MPDER); and \textit{Socially Complex Emotion Analysis}, focusing on Humor Understanding (HU) and Sarcasm Detection (SD).

\noindent \textbf{Evaluation Metrics.}
(1) EMER: Following~\cite{emotion-llama}, ChatGPT~\cite{gpt4o} rates ``Clue/Label Overlap'' (scale 0-10); 
(2) SABER-Test: ChatGPT scores responses (scale 0-3) across six dimensions: Video Description (VD), Speech Content Description(SCD), Acoustic Features Description (AFD), Facial Expression (FE), Body Language (BL), and Comprehensive Reasoning (CE); 
(3) EmoBench-M: We report the average Accuracy (ACC).

\noindent \textbf{Comparison Models.} We benchmark against three SOTA categories: (1) Omni Models: Qwen2.5-Omni (3B/7B)~\cite{Qwen2.5-Omni}, Qwen3-Omni-30B~\cite{qwen3-omni}, and Intern-s1-9B~\cite{intern-s1}; (2) Multimodal LLMs: generic (e.g., InternVL2.5-8B~\cite{internvl}, Video-LLaMA-7B~\cite{VideoLLaMA}, LongVA-DPO-7B~\cite{longva}, VideoChatGPT~\cite{video-chatgpt}, PandaGPT~\cite{pandagpt}, Valley~\cite{valley}) and emotion-specialized (Emotion-LLaMA-7B~\cite{emotion-llama}, EmoChat-7B~\cite{emochat}); and (3) Large-scale Models: Gemini-2.5-Pro~\cite{gemini2.5}/2.0-Flash~\cite{gemini2.0}, GLM-4V-Plus~\cite{glm4}, and Qwen2.5-VL-72B~\cite{qwen2.5vl}/InternVL2.5-78B~\cite{internvl}.

\noindent \textbf{Implementation Details.} We adopt Qwen2.5-Omni-7B as the base model. All experiments are conducted on 8 NVIDIA A100 GPUs. The training pipeline consists of two stages: first, an SED-based supervised fine-tuning stage trained for 2 epochs with a batch size of 128 and a learning rate of $1 \times 10^{-4}$, followed by a CA-DPO stage trained for 1 epoch with a batch size of 64 and a learning rate of $1 \times 10^{-5}$. To construct preference pairs for DPO, we sample $k$=10 responses per query at a temperature of 0.8 to select high-quality positive and negative pairs.

\begin{table}[t]
\centering
\caption{Performance comparison on the EMER. The best results are \textbf{bolded} and the second-best are \underline{underlined}.}
\label{tab:emer_results}
{
\resizebox{0.8\linewidth}{!}{
\begin{tabular}{l c c}
\toprule
\textbf{Model} & \textbf{Clue Overlap}↑ & \textbf{Label Overlap}↑ \\
\midrule
VideoChat-Text-7B~\cite{video-chat}  & 6.42 & 3.94 \\
Video-LLaMA-7B~\cite{VideoLLaMA}     & 6.64 & 4.89 \\
Video-ChatGPT-7B~\cite{video-chatgpt}   & 6.95 & 5.74 \\
PandaGPT-7B~\cite{pandagpt}        & 7.14 & 5.51 \\
VideoChat-Embed-7B~\cite{video-chat} & 7.15 & 5.65 \\
Valley-7B~\cite{valley}          & 7.24 & 5.77 \\
Emotion-LLaMA-7B~\cite{emotion-llama}   & 7.83 & 6.25 \\
EmoChat-7B~\cite{emochat}         & \underline{8.17} & 6.83 \\
Intern-S1-9B~\cite{intern-s1}    & 6.17 & 6.13 \\
Qwen2.5-Omni-7B~\cite{Qwen2.5-Omni} & 7.18 & 6.54 \\
Qwen3-Omni-30B~\cite{qwen3-omni}  & 7.83 & 6.86 \\
Gemini-2.5-Pro~\cite{gemini2.5}  & 8.07 & \textbf{7.08} \\
\midrule
\textbf{SABER-LLM}   & \textbf{8.25} & \underline{7.02} \\
\bottomrule
\end{tabular}}
}
\vspace{-12pt}
\end{table}

\begin{table}[t]
\centering
\caption{Performance comparison on the \textbf{SABER-Test} Consistent and Inconsistent subsets. The best results are highlighted in \textbf{bold} and the second-best results are \underline{underlined}.}
\label{tab:model_comparison}
% 使用 resizebox 确保表格适应页面宽度
\resizebox{\linewidth}{!}{
\begin{tabular}{lcccccccccccc}
\toprule
\multirow{2}{*}{\textbf{Model}} & \multicolumn{6}{c}{\textbf{Inconsistent}} & \multicolumn{6}{c}{\textbf{Consistent}} \\
\cmidrule(lr){2-7} \cmidrule(lr){8-13}
 & VD & SCD & AFD & FE & BL & CE & VD & SCD & AFD & FE & BL & CE \\
\midrule
Qwen2.5-Omni-3B      & 1.66 & 2.28 & 0.46 & 0.83 & 1.08 & 1.06 & 1.77 & 2.38 & 0.62 & 1.08 & 1.13 & 1.32 \\
Qwen2.5-Omni-7B      & 1.75 & 2.30 & 0.73 & 1.02 & 1.29 & 1.32 & 1.86 & 2.38 & 0.90 & 1.29 & 1.42 & 1.54 \\
Intern-s1-9B         & 1.95 & 1.03 & 0.65 & 1.44 & 1.76 & 1.44 & 2.09 & 1.02 & 0.89 & 1.54 & 1.85 & 1.58 \\
Qwen3-Omni-30B       & 2.31 & 2.45 & 2.27 & 1.95 & 2.17 & 1.96 & \underline{2.55} & 2.61 & 2.54 & 2.37 & 2.42 & 2.44 \\
SABER-LLM-7B     & \underline{2.43} & \textbf{2.72} & \textbf{2.65} & \underline{2.45} & \underline{2.45} & \underline{2.54} & 2.47 & \textbf{2.72} & \textbf{2.71} & \underline{2.60} & \underline{2.48} & \underline{2.67} \\
\midrule
Gemini-2.5-Pro-128B  & \textbf{2.65} & 2.63 & 2.57 & \textbf{2.56} & \textbf{2.55} & \textbf{2.59} & \textbf{2.63} & 2.60 & 2.60 & \textbf{2.60} & \textbf{2.56} & \textbf{2.68} \\
\bottomrule
\end{tabular}
}
\vspace{-13pt}
\end{table}

\begin{table}[t]
\centering
\caption{Ablation and Scalability Analysis on SABER-Test. We investigate the effectiveness of the SED paradigm across model scales and the contribution of CA-DPO. Best results are \textbf{bolded}.}
\label{tab:ablation}
\resizebox{\linewidth}{!}{
\begin{tabular}{l cccccc cccccc}
\toprule
\multirow{2}{*}{\textbf{Model}} & \multicolumn{6}{c}{\textbf{Inconsistent}} & \multicolumn{6}{c}{\textbf{Consistent}} \\
\cmidrule(lr){2-7} \cmidrule(lr){8-13}
 & VD & SCD & AFD & FE & BL & CE & VD & SCD & AFD & FE & BL & CE \\
\midrule
% --- 3B Comparison ---
Qwen2.5-Omni-3B (Base) & 1.66 & 2.28 & 0.46 & 0.83 & 1.08 & 1.06 & 1.77 & 2.38 & 0.62 & 1.08 & 1.13 & 1.32 \\
\hspace{0.5em} w/ SED & 2.29 & 2.65 & 2.57 & 2.35 & 2.37 & 2.41 & 2.37 & 2.63 & 2.64 & 2.47 & 2.64 & 2.58 \\
\midrule
% --- 7B Comparison ---
Qwen2.5-Omni-7B (Base)& 1.75 & 2.30 & 0.73 & 1.02 & 1.29 & 1.32 & 1.86 & 2.38 & 0.90 & 1.29 & 1.42 & 1.54 \\
\hspace{0.5em} w/ SED  & 2.36 & 2.67 & 2.59 & 2.39 & 2.36 & 2.49 & 2.44 & 2.69 & 2.69 & 2.56 & 2.44 & 2.64 \\
\hspace{0.5em} w/ SED + CA-DPO & \textbf{2.43} & \textbf{2.72} & \textbf{2.65} & \textbf{2.45} & \textbf{2.45} & \textbf{2.54} & \textbf{2.47} & \textbf{2.72} & \textbf{2.71} & \textbf{2.60} & \textbf{2.48} & \textbf{2.67} \\
\bottomrule
\end{tabular}
}
\vspace{-10pt}
\end{table}

\vspace{-5pt}
\subsection{Comparison with State-of-the-Art Methods}
\vspace{-3pt}
We evaluate SABER-LLM on two benchmarks. On EmoBench-M (Table~\ref{tab:emobench}), SABER-LLM (7B) establishes a new open-source standard with an average score of 59.88. Notably, it outperforms significantly larger architectures like Qwen3-Omni-30B, demonstrating superior parameter efficiency. Particularly in PEA and FCDEA, our model achieves scores of 76.15 and 73.96 respectively, approaching the closed-source Gemini-2.5-Pro.

% We attribute the performance gap on the SPER to the inherent modality competition within general-purpose MLLMs. While these models possess broad audio-visual comprehension, their joint embedding space is often dominated by high-level semantic or event-level features. Consequently, subtle paralinguistic signals (e.g., vocal tension, speech hesitations, or tonal shifts) are frequently suppressed or filtered out as noise during the end-to-end inference process.
% In contrast, SABER-LLM addresses this through the SED paradigm, which serves as a structural bottleneck for explicit grounding. By mandating the independent extraction of acoustic evidence before high-order reasoning, SED prevents these fragile affective cues from being overshadowed by more salient linguistic content. This mechanism ensures that paralinguistic nuances are preserved as primary evidence that the model must reconcile, effectively resolving the misalignment between surface-level perception and deep affective reasoning.
We attribute the performance gap on SPER to the inherent modality competition within general-purpose MLLMs. While these models possess broad comprehension, their joint embedding space is often dominated by high-level semantic features, leading subtle paralinguistic signals (e.g., vocal tension, speech hesitations, or tonal shifts) to be suppressed as noise during end-to-end inference.
SABER-LLM addresses this through the SED paradigm, which serves as a structural bottleneck for explicit grounding. By mandating the independent extraction of acoustic evidence before reasoning, SED prevents fragile affective cues from being overshadowed by more salient linguistic content. This ensures that paralinguistic nuances are preserved as primary evidence for the model to reconcile, effectively resolving the misalignment between surface perception and deep affective reasoning.

On EMER dataset, as shown in Table~\ref{tab:emer_results}, SABER-LLM achieves a Clue Overlap score of 8.25, surpassing all open-source baselines and even Gemini-2.5-Pro. In terms of Label Overlap score, it ranks second only to Gemini-2.5-Pro.

\vspace{-6pt}
\subsection{Robustness against Modal Inconsistency}
\vspace{-3pt}
We analyze performance discrepancies between inconsistent and consistent subsets to probe model's stability in complex scenarios. As shown in Table \ref{tab:model_comparison}, baseline models struggle due to uni-modal dominance. In inconsistent scenarios, AFD score of Qwen2.5-Omni-7B drops from 0.90 to 0.73. This decline exposes the vulnerability of existing models that when visual signals are dominant, they struggle to maintain independent perception of auditory cues. Similarly, FE score drops from 1.29 to 1.02, indicating the model captures only the superficial emotion while overlooking deeper, contradictory micro-expressions.
In contrast, SABER-LLM maintains robustness. In the challenging inconsistent subset, our model achieves an AFD score of 2.65, surpassing even the closed-source Gemini-2.5-Pro. This result validates that the SED strategy effectively decouples modal dependencies, ensuring the model can accurately capture subtle acoustic features and visual anomalies even under strong cross-modal distraction.

\begin{figure*}[]
  \centering
  \includegraphics[width=0.83\linewidth]{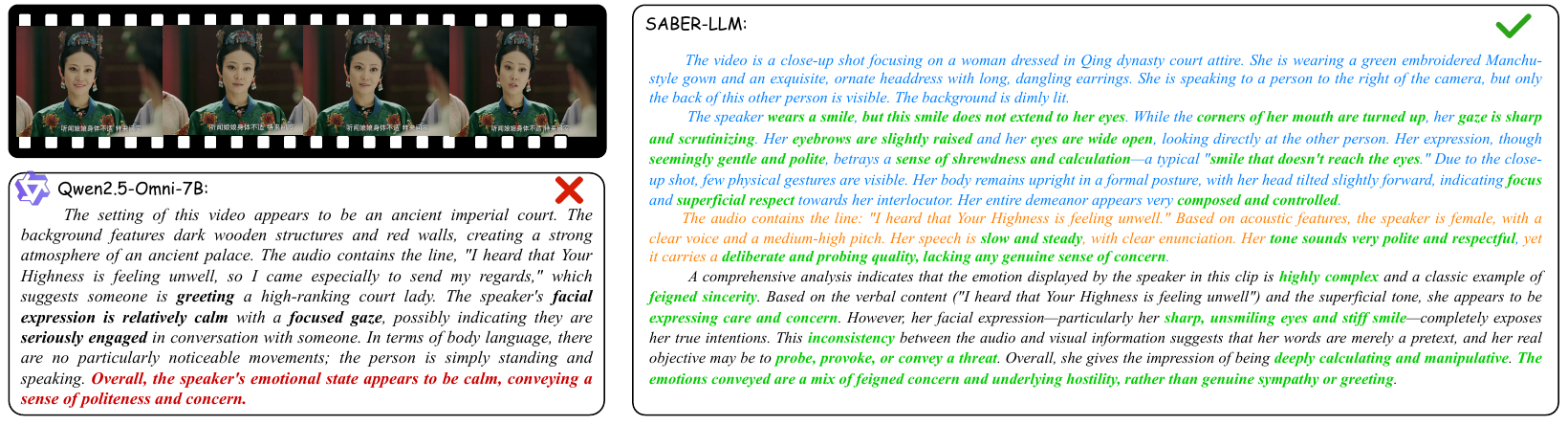}  % 使用单栏宽度的80%
  \caption{Qualitative comparison on a 'Feigned Sincerity' case. While the baseline is misled by the visual smile into predicting "Concern," SABER-LLM correctly identifies the mismatch between the "unsmiling eyes" and the "probing tone," successfully deducing the underlying manipulative intent.}
  \label{fig:qualitative_case}
  \vspace{-12pt}
\end{figure*}

\vspace{-5pt}
\subsection{Ablation Study}
\label{sec:ablation}
\vspace{-5pt}
% (1) Cross-Scale Effectiveness of SED: Comparing models at the 3B scale, the baseline performs poorly in acoustic perception under inconsistent scenarios (scoring only 0.46). However, the introduction of the SED paradigm in Qwen-Omni-3B + SED substantially boosts this score to 2.57. This result indicates that the improvement stems from the ``perceive-then-reason'' methodology rather than mere parameter scaling.
Table~\ref{tab:ablation} verifies the effectiveness and scalability of our core components. 
(1) Effectiveness of SED across scales: we observe that the 3B parameters model struggles to decode acoustic cues when they contradict other modalities. The introduction of SED yields a significant performance increase from 0.46 to 2.57, nearly a five-fold increase. This pronounced improvement underscores that robust multimodal reasoning is driven by the rigorous disentanglement of evidence facilitated by SED, independent of model size.

(2) Refinement via DPO: CA-DPO further impove AFD score of our model in inconsistent scenarios from 2.59 to 2.65 and further improves the comprehensive reasoning score. 
This confirms that the alignment strategy effectively refines the model's decision boundaries, enabling it to better suppress hallucinations and enhance comprehensive reasoning in contradictory samples.

\vspace{-5pt}
\subsection{Qualitative Analysis}
\label{sec:qualitative_analysis}
\vspace{-2pt}
Figure~\ref{fig:qualitative_case} presents a ``Feigned Sincerity'' case to demonstrate reasoning depth. The speaker delivers a verbally respectful message with a smile, yet harbors underlying hostility. Qwen2.5-Omni fails due to unimodal dominance, misled by the visual ``smile'' and polite text into predicting ``Concern.'' Conversely, leveraging structured reasoning, SABER-LLM decouples the visual anomaly (``smile doesn't reach eyes'') and acoustic nuance (``probing quality''). Crucially, SABER-LLM does not merely fuse modalities; it resolves cross-modal conflicts. While baselines fall for the visual 'trap' (the smile), our model explicitly flags the acoustic mismatch ('probing quality'), effectively vetoing the superficial positive prediction.

\vspace{-5pt}
\section{Conclusion}
\label{sec:conclusion}
\vspace{-5pt}
We proposed SABER-LLM to address ``uni-modal dominance'' caused by coarse-grained perception. We introduced SABER, a large-scale dataset annotated with a six-dimensional schema to support deep causal reasoning. Furthermore, our SED and CA-DPO enforce a ``perceive-then-reason'' paradigm, ensuring conclusions are based on disentangled evidence. Experiments on EmoBench-M, EMER, and SABER-Test demonstrate that SABER-LLM outperforms existing open-source models and achieves robustness competitive with other large-scale and closed-source models.

\vspace{-5pt}

\bibliographystyle{IEEEbib}
\bibliography{icme2026references}
\end{document}